# Fostering Joint Innovation: A Global Online Platform for Ideas Sharing and Collaboration

Hossein Jamali, Sergiu M. Dascalu, Frederick C. Harris, Jr.
*Department of Computer Science and Engineering*
*University of Nevada, Reno*
Reno, NV, USA
Email: hossein.jamali@nevada.unr.edu, {dascalus, fred.harris}@cse.unr.edu

*Abstract*—In today's world, where moving forward hinges on innovation and working together, this article introduces a new global online platform that's all about sparking teamwork to come up with new ideas. This platform goes beyond borders and barriers between different fields, creating an exciting space where people from all over the world can swap ideas, get helpful feedback, and team up on exciting projects. What sets our platform apart is its ability to tap into the combined brainpower of a diverse bunch of users, giving people the power to come up with game-changing ideas that tackle big global problems. By making it easy for people to share ideas and promoting a culture of working together, our platform is like a buddy for innovation, boosting creativity and problem-solving on a global level. This article spills the details on what the platform aims to do, how it works, and what makes it special, emphasizing how it can kickstart creativity, ramp up problem-solving skills, and get different fields collaborating. It's not just a tool – it's a whole new way of teaming up to make daily life better and build a global community of problem-solving pals.

*Index Terms*—Social networking, Innovation, Knowledge sharing, Collaboration Space, Global Online Platform.

## I. INTRODUCTION

In the contemporary world, where interconnectivity is the norm, conventional methods of generating ideas and problem-solving often encounter substantial obstacles. These challenges are frequently due to limited perspectives, a lack of collaboration, and geographical restrictions. Current platforms for sharing ideas and collaboration are fragmented and specialized, thereby limiting cross-disciplinary interactions and the diversity of viewpoints necessary for effective problem-solving. To tackle these issues, our project envisions a global online platform dedicated to collaborative idea sharing and development.

During the project's discovery phase, research focused on understanding user needs and expectations through surveys, interviews, and feedback. The platform's design emphasizes simplicity, accessibility, and interactivity to facilitate easy idea submission, efficient feedback exchange, and seamless networking opportunities. To overcome existing limitations, our solution incorporates advanced recommendation systems using machine learning algorithms to suggest relevant ideas and potential collaborators, encouraging interdisciplinary collaborations and exploring new domains.

The prototype design pays careful attention to privacy and intellectual property protection, allowing users control over the visibility of their ideas. This ensures a secure environment for freely expressing and developing ideas without concerns about theft or misuse. Overall, our project's solution offers a comprehensive platform to address the challenges of traditional idea-generation methods. By providing a global space for collaboration, feedback, and networking, we aim to foster a community of diverse problem solvers, accelerate innovation, and create tangible solutions to global challenges. The project's concept, discovery, and design stages lay a strong foundation for developing a transformative platform empowering individuals to share ideas, collaborate effectively, and make a positive impact on society.

## II. PREVIOUS WORKS

Collaboration and innovation are pivotal for addressing complex challenges. Our proposed global online platform catalyzes collaborative innovation, enabling users to share ideas and engage creatively. Platforms like Online Base Innovation (OBI) have supported globally distributed learning, emphasizing collaborative concept development [1]. Interdisciplinary courses enhance science communication and collaboration, fostering novel educational materials and approaches [2].

Combining collaboration and competition in idea management can lead to improved outcomes. Recent research suggests that these two approaches can complement each other. However, it's crucial to address intrinsic motivation, facilitate knowledge sharing, and consider rewards and company culture [3].

Collaboration mechanisms have been employed in multinational firms' internal idea management, showcasing effective coexistence of collaboration and competition [3]. In the realm of collaborative ideation, managing appropriate mechanisms, such as patents, copyrights, legal agreements, and secrecy, is essential for success. The timing of applying each mechanism remains a challenge.

The cons of collaboration include increased coordination costs and the necessity to share potential rewards of successful ideas. While working with a team provides access to additional resources, it can deter idea generators due to increased coordination costs. Determining the right time to apply appropriability mechanisms and fostering safe knowledge exchange is essential for effective collaborative ideation [4], [5].

## III. COLLABORATIVE INNOVATION: CONCEPTUAL FRAMEWORK

### A. Importance of Collaborative Innovation

Collaborative innovation is a key driver of success, fostering creativity and problem-solving through collective ambition. In today's global economy, it unites stakeholders to enhance innovation, address challenges, and promote sustainable development [7], [8], [9]. This approach enables effective responses to changing demands and encourages knowledge sharing, learning, and skill development for long-term growth [10]. The power of collaborative innovation is evident in interdisciplinary fields, such as the study by Bazmara, Mianroodi, and Silani in 2023, showcasing successful applications like physics-informed neural networks for complex problem-solving and emphasizing the benefits of diverse collaboration [11].

### B. Mechanisms and Pathways for Deepening Collaborative Innovation

To deepen collaborative innovation, organizations should design mechanisms for resource, benefits, and management coordination. Strategies include increased investment, promoting the triple-helix model, establishing decision-making mechanisms, and optimizing collaboration by fostering complementarity. Scientific benefit distribution, strengthened evaluation feedback, and improved collaborative environments further enhance the coordination mechanism, unlocking collaborative innovation's full potential for sustainable growth and development [10].

## IV. PLATFORM OBJECTIVES

### 1) Fostering Creativity, Innovation, and Collaboration

With an easy-to-use interface, users can easily share ideas, get feedback, and actively engage in the development and refinement of new ideas in an environment that encourages active involvement and diversity of views.

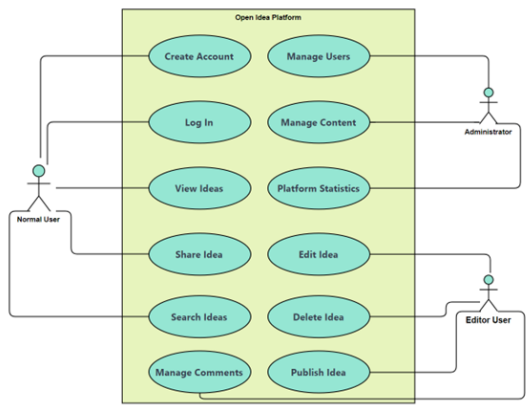

Fig. 1: The Use Case Diagram for Idea Sharing Platform

**Use Case Diagram (Fig. 1):** Figure 1 shows the main interactions and functions of our idea sharing platform. It shows different players and their specific roles in the system. Designed to encourage creativity, innovation, and collaboration, Figure 1 shows the structure of the idea sharing platform.

```
// Pseudo-code of the Idea Sharing Platform
function IdeaSharingPlatform():
    // Initialize the Idea Sharing platform
    InitializePlatform()

    while not EndConditionSatisfied():
        // Retrieve user's action or input
        UserAction = GetUserAction()

        if UserAction == "ViewIdeas":
            DisplayIdeas()

        else if UserAction == "ShareIdea":
            SubmitIdea()
            if IdeaIsValid():
                PublishIdea()
            else:
                RejectIdea()

    // Other user actions...

    // Update the state of the Idea
    // Sharing platform
    UpdatePlatformState()

    // Return the best idea based on user
    // ratings and feedback
    return BestIdea
```

Fig. 2: Pseudo-code of the Idea Sharing Platform is the main functionality

**Pseudo-code (Fig. 2):** Figure 2 presents the pseudo-code, emphasizing the platform's functionality in promoting creativity, innovation, and collaboration among users with diverse backgrounds. The user-friendly interface facilitates the exchange of ideas and dynamic feedback, serving as a fundamental element in achieving the platform's core objective of encouraging idea development, refinement, and collective contribution from diverse perspectives.

### 2) Enhancing Quality of Life

Our platform aims to improve the quality of life by leveraging user collective intelligence and diverse perspectives. Through collaboration and idea-sharing, we strive to address global challenges and develop innovative solutions with a positive societal impact. An example is the collaborative work of Paylakhi et al. [12], showcasing transformative problem-solving. By providing a space for diverse individuals to share ideas and collaborate, our platform contributes to advancements in healthcare, education, sustainability, technology, and more, stimulating the generation of transformative solutions for pressing global challenges.

*3) Empowering Users to Make a Difference*

Our platform endeavors to empower our users by respecting their ideas and allowing them to make a positive impact. Diverse users joining together encourage interdisciplinary research. The comparative study of Decision Tree, AdaBoost, Random Forest, Naïve Bayes, KNN, and perceptron for heart disease prediction [13] exemplifies this sort of endeavor. It shows how different perspectives can work to solve medical problems. Our platform seeks to encourage a collaborative spirit among researchers in fields across the spectrum, creating an international community of problem solvers.

*4) Creating a Global Community of Problem Solvers*

Our platform's goal is to create a space for sharing information and connecting people who have similar ideas to collaborate. It will stimulate feelings of belonging while encouraging everyone to become an active participant in society in these times. In this case, the highlighting of existing interdisciplinary activities is of great significance. For example, the work of Roshanzamir demonstrates how machine learning can be used in the field of transportation [14]. These include Mokhtari et al.'s study that surprises people by bringing together mathematics and chemistry [15], demonstrating how in interdisciplinary action outstanding presentations are produced. In promoting such an atmosphere, we will be fortunate if like-minded souls from around the world can band together to solve the problems with multi-faceted and comprehensive solutions.

*5) Fostering Lifelong Learning*

In addition to promoting creativity and innovation, our platform encourages lifelong learning and personal growth. Engaging in discussions, receiving feedback, and collaborating with others provide users the opportunity to expand their knowledge, develop new skills, and foster innovation within a thriving global community.

## V. PLATFORM FUNCTIONALITY AND CHARACTERISTICS

*1) Idea Sharing and Feedback*

The platform allows users to easily express ideas in a great variety of forms, thus producing a stimulating creative space. As we can see in Figure 3, the concise and clean interface makes effective conversations possible, letting us share ideas with one another through these channels as comments and feedback.

*2) Rating and Commenting System*

The platform incorporates a system for users to collectively evaluate and rate ideas based on relevance, feasibility, originality, and potential impact. This process helps surface the most promising concepts, and users can provide constructive comments to offer insights and alternative perspectives for idea enhancement.

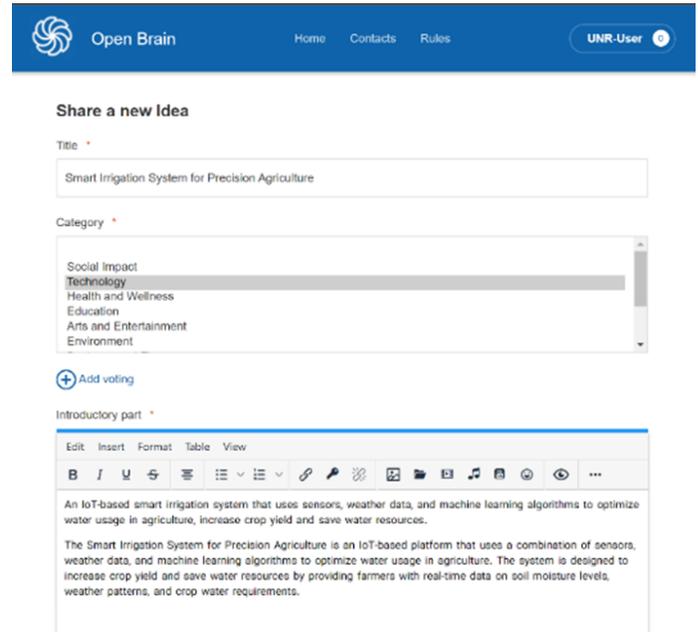

Fig. 3: Idea Sharing user interface.

*3) Networking and Collaboration Space*

As a dynamic, interconnected hub, the platform links users of like-minded interests. This makes teamwork more effective, encouraging innovation in what ideas are developed and how they manifest themselves. They can coordinate efforts and augment each other's small ideas with a multitude of new materials and suggestions while sharing the pleasant burden of bringing larger projects before more people who will be able to see them.

*4) Team Formation and Project Management Tools*

The platform facilitates collaborative efforts by offering tools for team formation and project management. Users can create or join project groups dedicated to specific ideas, enabling task assignment, deadline setting, and progress tracking. This centralized space streamlines collaboration, ensuring effective communication among team members, regardless of their geographical locations.

*5) Privacy and Intellectual Property Protection*

Privacy and intellectual property are important to us. Users retain control over the visibility and availability of their ideas and determine the scope of collaboration. Strong security measures are in place to prevent unauthorized access to user data. The platform promotes respect for intellectual property rights by encouraging users to share ideas responsibly and by providing clear attribution and sharing guidelines. We are also working on an information verification system, inspired by the strategies suggested by Yavary et al. [16]. This holistic approach not only sets the platform up as a safe space for collaboration but also works as a preventative tool against disinformation, building trust and innovation in a global community of trust.

*6) Knowledge Repository and Search Functionality*

The platform serves as a knowledge repository, aggregating ideas and collaborations. Its efficient search feature allows users to explore existing ideas and projects, preventing duplication of efforts and encouraging interdisciplinary collaborations.

*7) Gamification and Incentive Mechanisms*

In our platform, active users are recognized and rewarded for their valuable contributions through gamification and incentives.

*8) Users Group Policy*

Illustrated in Figure 4, our user group policy establishes a hierarchical framework for user roles, including Administrators, Chief Editors, Editors, Visitors, and Guest users, enabling efficient user management with customizable access levels.

Fig. 4: Users group policy.

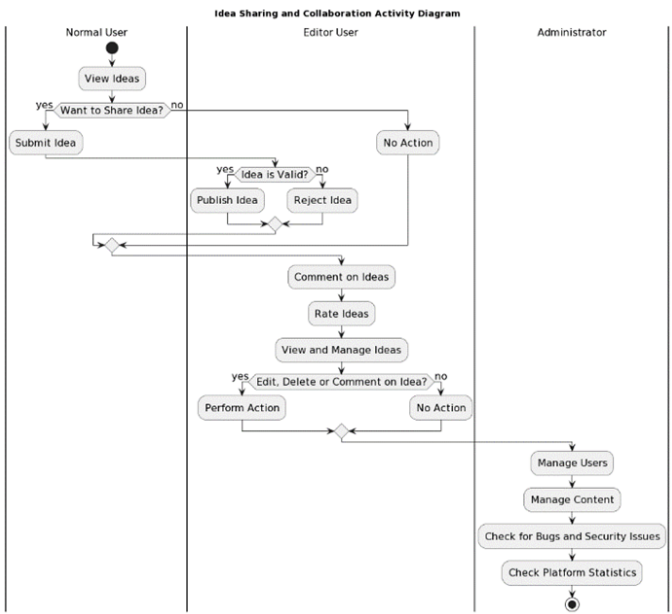

Fig. 5: The Idea Sharing Platform Activity Diagram.

The flexibility of the platform allows dynamic adaptation of the user group policy, ensuring scalability and tailored user experience. The associated activity diagram in Figure 5 provides a comprehensive overview of the Idea Sharing project's workflow, depicting key activities and interactions. Users seamlessly engage in idea submission, review, and publication, fostering a constructive feedback ecosystem. The diagram emphasizes networking, collaboration spaces, and administrative roles, encapsulating the core functionalities of the Idea Sharing platform.

## VI. COMPARISON AND EVALUATION

The platform proposed in this study was fully implemented in the Software Systems Lab at UNR. The implementation utilized the PHP programming language, JavaScript, and an SQL database on a Linux web server.

To evaluate the characteristics of our platform, we compared its features with those of similar platforms, including IdeaScale, Spigit, Brightidea, HYPE Innovation, and Mindjet. Two key aspects were considered in this comparison: Fragmentation and Idea Generation. Our observations are presented in Tables I and II below.

| Aspect | Fragmentation |
| --- | --- |
| Proposed Idea | Not fragmented, global online platform. |
| IdeaScale | Can be fragmented, catering to specific industries or communities [17], [18]. |
| Spigit | Can be fragmented, catering to specific industries or communities [19]. |
| Brightidea | Can be fragmented, catering to specific industries or communities [20]. |
| HYPE Innovation | Can be fragmented, catering to specific industries or communities [21]. |
| Mindjet | Not fragmented, but primarily a mind-mapping and project management tool [22]. |

TABLE I: Fragmentation

| Aspect | Idea Generation |
| --- | --- |
| Proposed Idea | Global space for collaboration, feedback, and networking. |
| IdeaScale | Cloud-based platform for capturing ideas and feedback [17], [18]. |
| Spigit | Software platform for crowdsourcing ideas and engaging employees, customers, and partners [19]. |
| Brightidea | Cloud-based platform for harnessing the collective intelligence of employees, customers, and partners [20]. |
| HYPE Innovation | Software platform for managing the innovation process [21]. |
| Mindjet | Software platform for capturing ideas, organizing information, and collaborating on projects [22]. |

TABLE II: Idea Generation

## VII. FUTURE WORK

The impact of our global online platform for collaborative innovation is immense, and continuous enhancements can amplify its impact. This includes continuous improvements to the user experience (UX) and user interface (UI), integration of user feedback to streamline navigation and interactions. Further enhancements to recommendation systems, virtual collaboration tools (VOCs), analytics integration, partnering with similar organizations, and continuous collection of user feedback will help build a platform that leverages creativity, collaboration, and problem-solving to make a real difference in society.

## VIII. CONCLUSION

Our proposed global collaborative innovation platform is a powerful game-changer, revolutionizing problem-solving and idea generation by bringing together experts and enthusiasts from all walks of life. We believe in creating a global community of innovators, unleashing the power of collective intelligence for mastering any challenge ahead. Our platform is an instrument for positive change, empowering individuals to drive growth and progress, step-by-step, one idea at a time.

## ACKNOWLEDGEMENTS

This material is based in part upon work supported by the National Science Foundation under grant OIA-2019609. Any opinions, findings, and conclusions or recommendations expressed in this material are those of the authors and do not necessarily reflect the views of the National Science Foundation.